\documentclass[twoside,a4paper,10pt]{article}

\usepackage[english,brazil]{babel}

\usepackage[utf8]{inputenc}
\usepackage[T1]{fontenc} 

\newcommand{\shorttitle}{Física na UFSM: uma análise de gênero dos egressos}

\usepackage{authblk} 
\usepackage{lipsum} 
\usepackage{times}
\usepackage{microtype} 
\usepackage[hmarginratio=1:1,top=33mm,columnsep=20pt,text={18cm,24.3cm}]{geometry} 
\usepackage{booktabs} 
\usepackage{abstract} 
\newcommand{\titleeng}{\vspace{4mm}\fontsize{14pt}{14pt}
\selectfont}
\newcommand{\titlept}{\textbf} 
\usepackage{titlesec} 
\usepackage{fancyhdr} 
\usepackage{amssymb,amsmath,amsthm,amsfonts} 
\usepackage{natbib} 
\usepackage{graphicx,rotating} 
\usepackage{enumerate}
\usepackage{graphics}
\usepackage[dvips]{epsfig}
\usepackage{multirow}
\usepackage{nonfloat}
\usepackage{multicol} 
\usepackage{url} 
\usepackage{hyperref} 
\usepackage{subfigure} 
\usepackage{bm}
\usepackage{color}
\usepackage{icomma} 

\allowdisplaybreaks[4] 

\title{
\titlept{O cinquentenário da física na UFSM: uma análise de gênero dos egressos} 
\\ 
\titleeng{The fiftieth anniversary of physics at UFSM: a gender analysis of graduates} 
} 

\author{Rogemar A. Riffel\footnote{Departamento de Física, UFSM}~  \& Lucio S. Dorneles$^*$} 
 
\date{}
 
\begin{document}

\maketitle 

\thispagestyle{empty} 

\begin{abstract}

\noindent A Universidade Federal de Santa Maria (UFSM) já formou mais de 500 licenciados, 135 bacharéis, 150 mestres e 90 doutores em física, tornando-se uma referência na área na região central do Rio Grande do Sul. Neste trabalho apresentamos uma análise do gênero dos egressos dos cursos de graduação e pós-graduação em física da UFSM. O percentual de egressos do gênero feminino tem aumentado com o tempo e é significativamente maior do que é observado nacionalmente em todos os cursos. Os valores percentuais médios de egressos do gênero feminino em relação ao total de egressos é de 35\,\%, 37\,\%,  28\,\%,  42 \,\% e 39 \,\% para os cursos de licenciatura diurno, licenciatura noturno, bacharelado, mestrado e doutorado, respectivamente. Observamos também um aumento da participação feminina entre egressos dos cursos de mestrado e doutorado em relação aos cursos de graduação, diferentemente do que se observa em números médios nacionais. Considerando-se somente os últimos 5 anos o percentual de egressos do gênero feminino atinge 50\,\% nos cursos de licenciatura diurno e mestrado em física. Embora este estudo seja de caráter local, ele reforça a necessidade da divulgação de dados referente ao gênero de egressos por todas as instituições de ensino superior, tornando possível a realização de estudos gerais.

\smallskip
\noindent \textbf{Palavras-chave:}  Física: egressos, Física: gênero.

\end{abstract}

{
\selectlanguage{english}
\begin{abstract}

\noindent The Federal University of Santa Maria (UFSM - {\it Universidade Federal de Santa Maria}) has trained more than 600 undergraduate and 200 Ph.D. students, becoming a reference in physics training at the Rio Grande do Sul central region. In this work we present a gender analysis of the UFSM graduated students from both levels. The number of graduated students of the female gender has been rising with time and is significatively bigger than the number observed nationally in both levels. The mean values are 35\,\%, 37\,\%,  28\,\%,  42 \,\% e 39 \,\% for the courses physics teaching (daytime), physics teaching (nightime), physics, master and doctorate, respectively. We also observe an increase in the participation of female students in graduate courses, when comparing to undergraduate courses, unlike what is observed in the national average numbers. Considering only the last 5 years the number of graduated students reaches 50\,\% in the physics teaching (daytime) and master courses. Although this study is of a local nature, it reinforces the need for the dissemination of data regarding the gender of graduates by all universities, allowing to conduct general gender studies.


\end{abstract}
}

\newpage


\section{Introdução}

A Universidade Federal de Santa Maria (UFSM) foi fundada em 1960. Em 1968 foi criado o Centro de Estudos Básicos na UFSM e juntamente com ele a formação em física é formalizada. No ano seguinte, ocorreu o ingresso da primeira turma do curso de Física -- Licenciatura Plena.  
 No princípio, o estudo de física na UFSM se restringia ao curso de licenciatura, com o objetivo de formar profissionais para atuar na educação básica. Em meados da década de 1990 foram fundados os cursos de física bacharelado e licenciatura plena noturno.  Deste a fundação do curso de física licenciatura no final dos anos 1960, a UFSM já formou  mais de 600 licenciados e bacharéis em física \citep{cfisica} e hoje é referência na formação de recursos humanos na área na região central do estado do Rio Grande do Sul. 

Parte significativa dos egressos dos cursos de física ingressam no Programa de pós-graduação em física da UFSM, que oferta o curso de mestrado desde 1994 e de doutorado desde 1999, tendo formado 145 mestres e 88 doutores em física até o final de 2016, distribuídos em três áreas de concentração: i) Áreas clássicas da fenomenologia e suas aplicações, ii) Astronomia e astrofísica e iii) Física da matéria condensada e seis linhas de pesquisa \citep{pgfisica}. Dessa forma, a UFSM tem um importante papel na formação de recursos humanos na área de física. 

\citet{bezerra16} apresentam um estudo sobre a questão de gênero na distribuição de bolsas em produtividade em pesquisa (PQ) do Conselho Nacional de Desenvolvimento Científico e Tecnológico (CNPq) no período de 2001 a 2011 e mostram que o percentual de mulheres bolsistas PQ é de somente 10\% e a participação feminina diminui do nível mais baixo (nível 2) para o nível mais alto (nível 1A). Já \citet{ferrari17} apresentam um estudo mais amplo, investigando além da questão de gênero a distribuição geográfica dos membros da academia brasileira de ciências.  Os autores encontram que somente 13\% dos membros titulares da academia brasileira de ciências são mulheres. Na área de física, este número cai para cerca de 7\%, sendo maior somente do que os percentuais vistos para as áreas de engenharias e matemática. Os autores também encontram que o percentual de mulheres bolsistas PQ diminui com o avanço na carreira, resultado similar ao apresentado em \citet{bezerra16}.  \citet{saitovich15} apresentam uma comparação entre participação feminina na área de física no Brasil com outras regiões do mundo e observam que o percentual de mulheres na área é menor do que 50\% em todas as regiões do mundo e em todos os níveis de ensino. Em nível de graduação, o valor percentual médio de mulheres é de cerca de 25\%, já em nível de doutorado este número diminui para pouco menos de 20\% e entre docentes a participação feminina é de somente 10\%. Para o Brasil, o percentual de mulheres na física é de 24\%, 16\% e 15\% na graduação, doutorado e carreira acadêmica, respectivamente \citep{saitovich15}. 

Apresentamos aqui um estudo estatístico do gênero dos egressos dos cursos de física e dos cursos de mestrado e doutorado do programa de pós-graduação em física da UFSM. Na seção \ref{dados} apresentamos a compilação dos dados e a metodologia adotada. A seção \ref{resultados} apresenta os resultados e discussões deste trabalho, abordando principalmente a questão de gênero na área de física na UFSM, os quais são comparados com a literatura na seção~\ref{disc}. Finalmente, as considerações finais são apresentadas na seção \ref{conc}.

\section{Metodologia}\label{dados}

\begin{table*}
\caption{Egressos dos cursos de graduação em física da UFSM, separados por gênero.}
\label{tab_g}
\begin{center}
\begin{tabular}{| c| c c c  |  c c c |  c c c|}
\hline
& \multicolumn{3}{|c|}{Física licenciatura diurno} & \multicolumn{3}{c|}{Física licenciatura noturno} & \multicolumn{3}{c|}{Física bacharelado} \\ 
Ano & Masculino & Feminino & Total  & Masculino & Feminino & Total & Masculino & Feminino & Total  \\ 
\hline
1972  &  3   &  0   &	  3         &	   &	  &	   &	     &  	&	 \\
1973  &  1   &  0   &	  1         &	   &	  &	   &	     &  	&	 \\
1974  &  1   &  1   &	  2         &	   &	  &	   &	     &  	&	 \\
1975  &  7   &  2   &	  9         &	   &	  &	   &	     &  	&	 \\
1976  &  7   &  2   &	  9         &	   &	  &	   &	     &  	&	 \\
1977  &  11  &  8   &	  19        &	   &	  &	   &	     &  	&	 \\
1978  &  3   &  7   &	  10        &	   &	  &	   &	     &  	&	 \\
1979  &  8   &  5   &	  13        &	   &	  &	   &	     &  	&	 \\
1980  &  1   &  0   &	  1         &	   &	  &	   &	     &  	&	 \\
1981  &  2   &  1   &	  3         &	   &	  &	   &	     &  	&	 \\
1982  &  5   &  1   &	  6         &	   &	  &	   &	     &  	&	 \\
1983  &  9   &  3   &	  12        &	   &	  &	   &	     &  	&	 \\
1984  &  9   &  2   &	  11        &	   &	  &	   &	     &  	&	 \\
1985  &  8   &  3   &	  11        &	   &	  &	   &	     &  	&	 \\
1986  &  6   &  6   &	  12        &	   &	  &	   &	     &  	&	 \\
1987  &  10  &  1   &	  11        &	   &	  &	   &	     &  	&	 \\
1988  &  6   &  1   &	  7         &	   &	  &	   &	     &  	&	 \\
1989  &  0   &  1   &	  1         &	   &	  &	   &	     &  	&	 \\
1990  &  8   &  3   &	  11        &	   &	  &	   &	     &  	&	 \\
1991  &  6   &  1   &	  7         &	   &	  &	   &	     &  	&	 \\
1992  &  6   &  4   &	  10        &	   &	  &	   &	     &  	&	 \\
1993  &  9   &  2   &	  11        &	   &	  &	   &	     &  	&	 \\
1994  &  3   &  1   &	  4         &	   &	  &	   &	     &  	&	 \\
1995  &  7   &  4   &	  11        &	   &	  &	   &	     &  	&	 \\
1996  &  11  &  1   &	  12        &	   &	  &	   &	     &  	&	 \\
1997  &  5   &  3   &	  8         &	   &	  &	   &	     &  	&	 \\
1998  &  5   &  3   &	  8         &	   &	  &	   &	2    &    2	&   4	 \\
1999  &  2   &  4   &	  6         &	   &	  &	   &	2    &    2	&   4	 \\
2000  &  5   &  2   &	  7         & 1    &	0 &    1   &	8    &    1	&   9	 \\
2001  &  7   &  7   &	  14        & 5    &	4 &    9   &	2    &    0	&   2	 \\
2002  &  2   &  2   &	  4         & 4    &	4 &    8   &	4    &    2	&   6	 \\
2003  &  0   &  3   &	  3         & 5    &	1 &    6   &	8    &    2	&   10   \\
2004  &  3   &  2   &	  5         & 7    &	6 &    13  &	3    &    4	&   7	 \\
2005  &  9   &  4   &	  13        & 10   &	6 &    16  &	7    &    0	&   7	 \\
2006  &  2   &  3   &	  5         & 8    &	1 &    9   &	6    &    1	&   7	 \\
2007  &  2   &  1   &	  3         & 5    &	0 &    5   &	6    &    1	&   7	 \\
2008  &  7   &  4   &	  11        & 5    &	2 &    7   &	7    &    4	&   11   \\
2009  &  5   &  1   &	  6         & 7    &	8 &    15  &	8    &    1	&   9	 \\
2010  &  5   &  4   &	  9         & 2    &	3 &    5   &	8    &    2	&   10   \\
2011  &  3   &  1   &	  4         & 4    &	0 &    4   &	2    &    0	&   2	 \\
2012  &  8   &  4   &	  12        & 3    &	4 &    7   &	4    &    4	&   8	 \\
2013  &  0   &  5   &	  5         & 5    &	1 &    6   &	1    &    1	&   2	 \\
2014  &  0   &  1   &	  1         & 3    &	1 &    4   &	2    &    3	&   5	 \\
2015  &  3   &  4   &	  7         & 3    &	4 &    7   &	5    &    0	&   5	 \\
2016  &  2   &  1   &	  3         & 1    &	1 &    2   &	4    &    5	&   9	 \\
\hline 
\end{tabular}
\end{center}
\end{table*}

\begin{table}
\label{tab_pg}
\caption{Egressos dos cursos de mestrado e doutorado em física da UFSM, separados por gênero.}
\begin{center}
\begin{tabular}{| c| c c c| c c c |}
\hline
& \multicolumn{3}{|c|}{mestrado} & \multicolumn{3}{c|}{doutorado} \\ 
Ano & Masculino & Feminino & Total & Masculino & Feminino & Total \\
\hline
1999  &    6	 &    1     &    7    & 	 & 	    & 		  \\
2000  &    2	 &    4     &    6    & 	 & 	    & 		  \\
2001  &    2	 &    4     &    6    &      1   & 	0   & 	   1	  \\
2002  &    3	 &    4     &    7    &      0   & 	0   & 	   0	   \\
2003  &    6	 &    2     &    8    &      1   & 	1   & 	   2	   \\
2004  &    4	 &    5     &    9    &      2   & 	1   & 	   3	   \\
2005  &    4	 &    2     &    6    &      2   & 	2   & 	   4	   \\
2006  &    7	 &    2     &    9    &      5   & 	4   & 	   9	   \\
2007  &    3	 &    5     &    8    &      3   & 	0   & 	   3	   \\
2008  &    6	 &    1     &    7    &      4   & 	2   & 	   6	    \\
2009  &    6	 &    2     &    8    &      5   & 	3   & 	   8	    \\
2010  &    5	 &    2     &    7    &      4   & 	1   & 	   5	    \\
2011  &    5	 &    2     &    7    &      3   & 	5   & 	   8	    \\
2012  &    2	 &    7     &    9    &      10  & 	2   & 	   12	    \\
2013  &    3	 &    3     &    6    &      4   & 	2   & 	   6	    \\
2014  &    4	 &    6     &    10   &      4   & 	3   & 	   7	     \\
2015  &    4	 &    2     &    6    &      4   & 	2   & 	   6	     \\
2016  &    4	 &    2     &    6    &      2   & 	6   & 	   8	     \\
\hline

\end{tabular}
\end{center}
\end{table}

Acessamos a {\it website} dos cursos de física da UFSM\footnote{http://www.ufsm.br/cfisica} \citep{cfisica} e compilamos o número de egressos nos cursos de física licenciatura plena diurno, física bacharelado e física licenciatura plena noturno em cada semestre. Considerando que o número de egressos por semestre em cada curso é pequeno e que o ingresso nos cursos se dá uma única vez por ano, separamos os egressos por gênero e calculamos o número total em períodos anuais para cada curso. Analisamos períodos desde a primeira turma de formandos até o ano de 2016. O curso de licenciatura em física diurno teve a primeira turma formada em 1972, já o curso de bacharelado teve sua primeira turma de egressos em 1998 e o curso de licenciatura noturno formou a primeira turma em 2000. 

Realizamos o mesmo procedimeno adotado para os cursos de graduação para os egressos do programa de pós-graduação em física da UFSM \citep{pgfisica}, cujas informações estão disponíveis na {\it website} do programa \footnote{http://www.ufsm.br/pgfisica}. 

A Tabela~\ref{tab_g} apresenta os dados compilados para os cursos de graduação em física, enquanto que a Tabela~\ref{tab_pg} mostra os números para os cursos de pós-graduação. Estes dados foram utilizados para a construção de gráficos do número total de egressos e percentual de egressos dos gêneros feminino e masculino.  Estes dados são comparados entre os diferentes cursos e níveis de ensino da UFSM, bem como com resultados disponíveis na literatura.

\section{Resultados}\label{resultados}

Utilizamos os dados coletados para a construção de gráficos do número total de egressos dos cursos de graduação, mestrado e doutorado em física e percentual de egressos de cada gênero vs. a data de formatura/defesa de dissertação ou tese, bem como para obtenção das notas médias dos egressos dos cursos de física em nível de graduação. 

\subsection{Uma análise de gênero dos egressos}

\begin{figure*}
\includegraphics[width=0.33\textwidth]{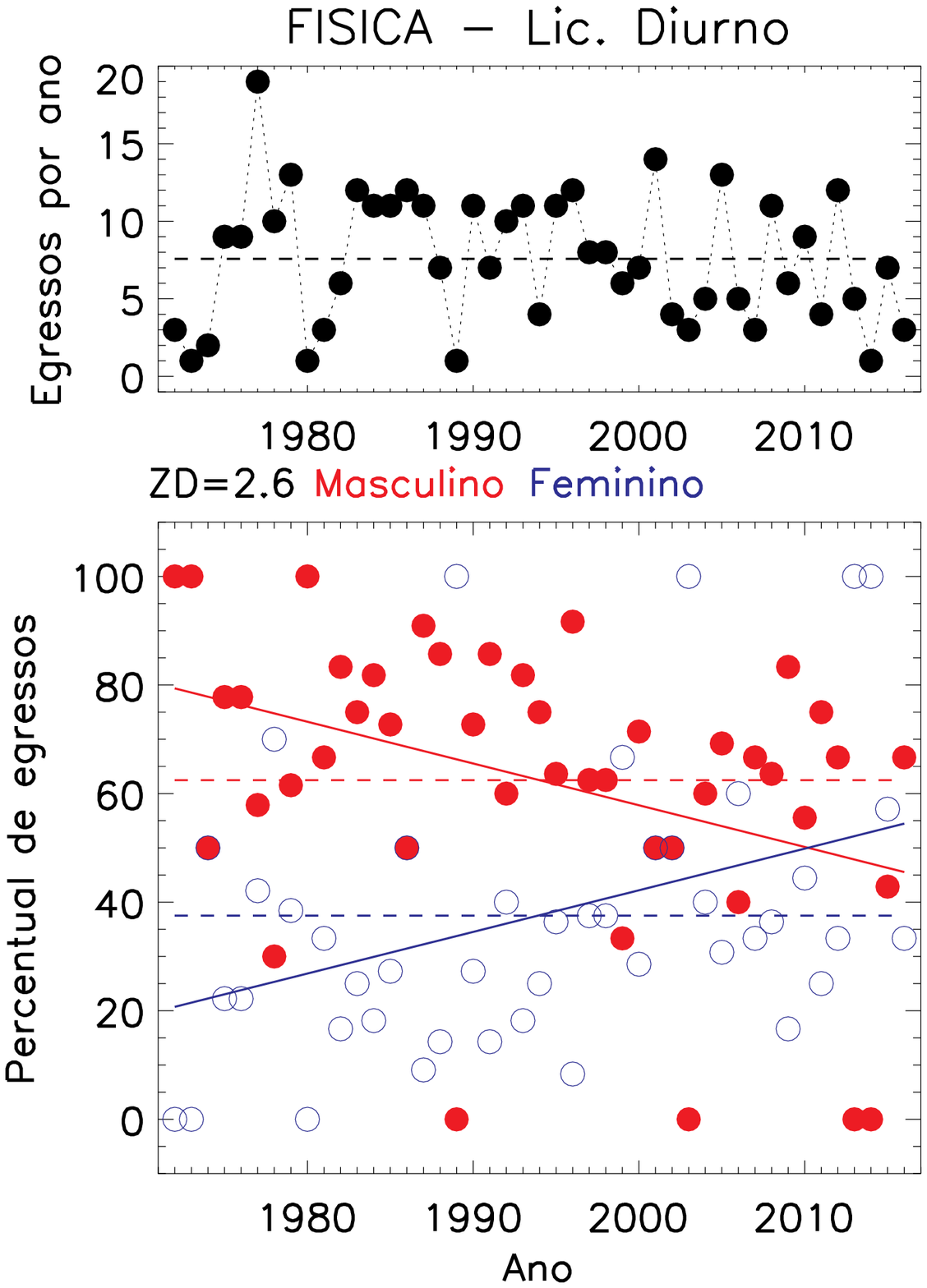}
\includegraphics[width=0.33\textwidth]{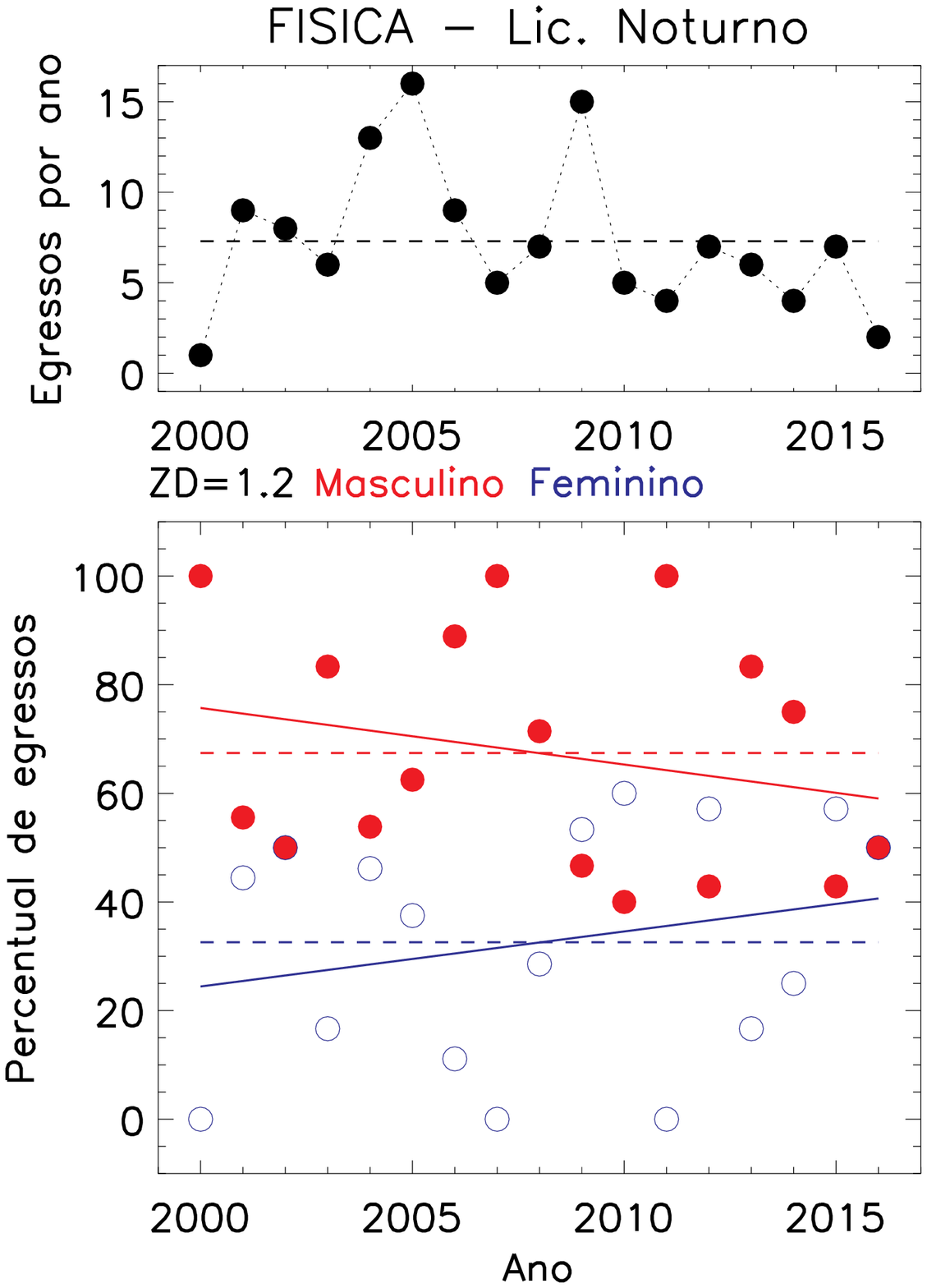}
\includegraphics[width=0.33\textwidth]{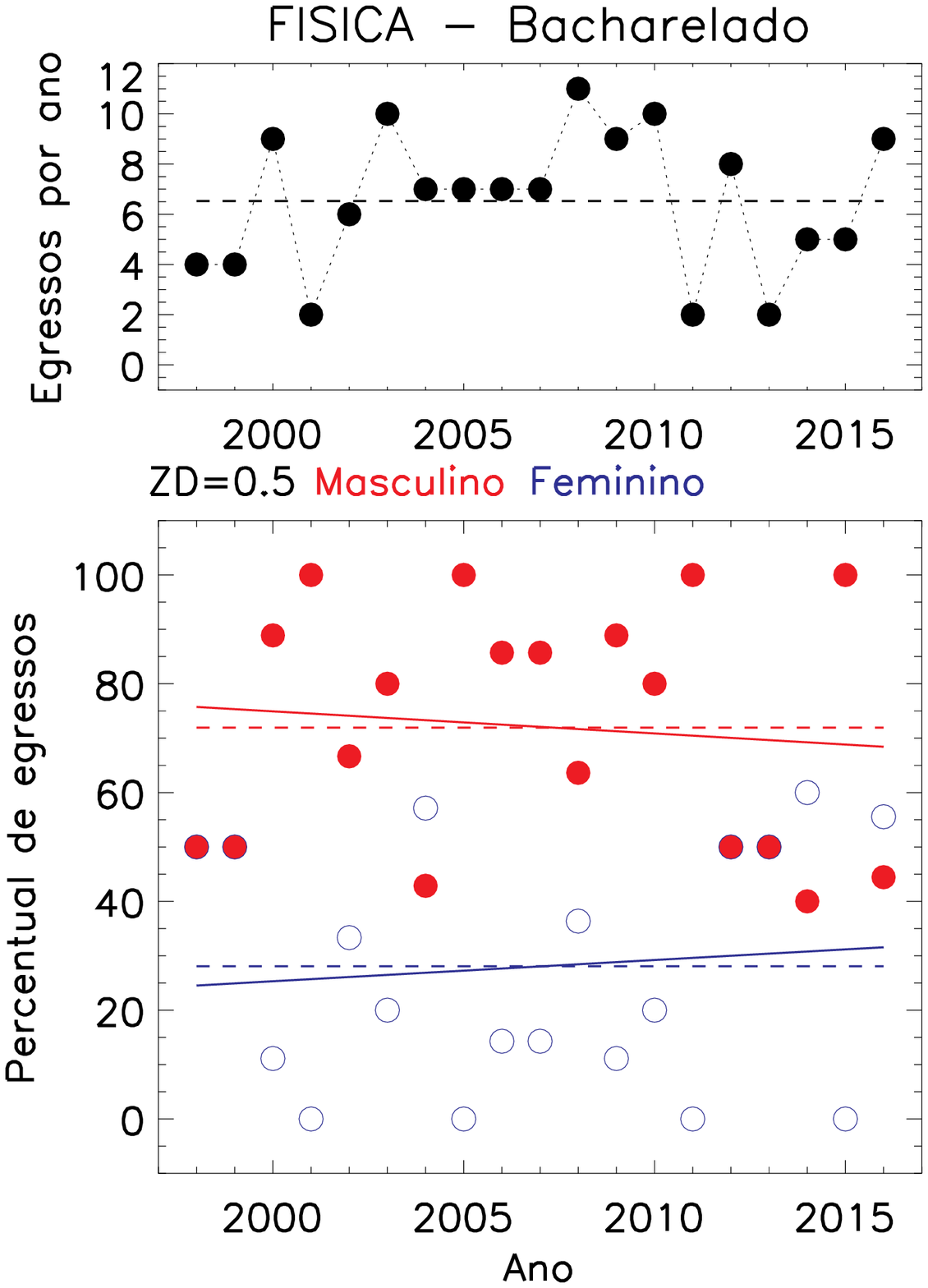}

\caption{Distribuição temporal do número total de egressos (topo) e percentual de egressos separados por gênero (base) para os cursos de física licenciatura plena diurno (esquerda), física licenciatura Plana noturno (centro) e física bacharelado (direita). Nos paineis inferiores, egressos do gênero masculino são representados por círculos vermelhos preenchidos e egressos do gênero feminino por circulos azuis vazados. Em todos os paineis as linhas tracejadas apresentam valores médios e as barras de incertezas correspondem ao erro padrão de cada distribuição. As linhas contínuas apresentadas nos paineis inferiores correspondem ao resultado de uma regressão linear de cada distribuição. No topo dos baineis inferiores apresenta-se o coeficiente $ZD$, que indica a significância estatística da correlação encontrada entre o percentual de egressos de cada gênero e a data de formatura.}
\label{graduacao}
\end{figure*}

A Figura~\ref{graduacao} apresenta os gráficos correspondentes para os diferentes cursos de graduação em física da UFSM. Os paineis da esquerda representam o curso de física licenciatura plena diurno, os paineis centrais são para o curso de física licenciatura plena noturno e o curso de física bacharelado é apresentado nos paineis da direita. Os paineis superiores mostram o número total de egressos em cada ano, desde a data da primeira turma formada em cada curso. A linha tracejada representa o valor médio do número de egressos em cada curso. O número médio de egressos no curso de física licenciatura diurno é de 7,6 graduados/ano; já para o curso de física licenciatura noturno a média do número de egressos é 7,3 graduados/ano, enquanto que o curso de física bacharelado forma em média 6,5 profissionais a cada ano. 

Os paineis inferiores da Figura~\ref{graduacao} mostram a distribuição do percentual de egressos de cada gênero em função do ano de formatura.  Nestas figuras, o gênero feminino é representado por símbolos e linhas azuis e o gênero masculino por símbolos e linhas vermelhas. Com o objetivo de verificar se existe correlação entre o percentual de egressos de cada gênero e ano de formatura, utilizamos a rotina  {\it r$_-$correlate.pro}, escrita na linguagem IDL ({\it Interactive Data Language})\footnote{$http://www.harrisgeospatial.com/SoftwareandTechnology/IDL.aspx$} para estimar a significância estatística ($ZD$) de uma possível correlação utilizando a estatística de Spermann. No topo de cada painel apresentamos o valor de $ZD$ obtido. Observa-se que para o curso de licenciatura diurno há uma correlação com significância acima de 2$\sigma$ ($ZD=2.6$), enquanto que para o curso de licenciatura noturno e bacharelado não se observa correlação, com $ZD=1.2$ e $0.5$, respectivamente. 

Considerando-se que observa-se uma correlação entre o percentual de egressos de cada gênero com o ano de formatura para o curso de física licenciatura diurno, decidimos ajustar as distribuições para cada gênero utilizando-se uma relação linear. Este ajuste foi realizado utlizando a rotina {\it linmix$_-$err.pro} que faz parte da biblioteca de rotinas IDL para uso em astronomia do {\it Goddard Space Flight Center - NASA}\footnote{Esta biblioteca está disponível em $https://idlastro.gsfc.nasa.gov/$}. Esta rotina adota uma abordagem Bayesiana para realizar a regressão linear, permitindo considerar incertezas em ambos os parâmetros. Considerando que o número de egressos varia  ano-a-ano, atribuímos incertezas poissionicas para o percentual de egressos, calculadas por $1/\sqrt{N_a}$, onde $N_a$ corresponde ao número de total egressos a cada ano.  Dessa forma leva-se em conta o tamanho da amostra durante a regressão linear. Os resultados do ajuste são apresentados nas figuras como linhas contínuas vermelhas e azuis para o genêro masculino e feminino, respectivamente.

Embora somente observe-se uma correlação entre o percentual de egressos de cada gênero e o ano de formatura para o curso de licenciatura diurno, as distribuições de pontos para a licenciatura noturno e bacharelado também apresentam uma tendência de aumento de egressos do gênero feminino com o passar do tempo. Observamos que $ZD=0.3$ para o curso de licenciatura diurno considerando apenas os egressos a partir do ano 2000. Os valores médios do percentual de egressos de cada gênero para os diferentes cursos de graduação são representados pelas linhas tracejadas nos paineis inferiores da Fig.~\ref{graduacao}. Para o curso de física licenciatura diurno o número médio de egressos do gênero feminino é de 35\,\%, calculado desde a formatura da primeira turma em 1972 até 2016. Para o curso de física licenciatura noturno, o valor médio é de 37\,\% e para o curso de física bacharelado o percentual médio de egressos do gênero feminino é de 28\,\%. 

\begin{figure*}
\centering
\includegraphics[width=0.4\textwidth]{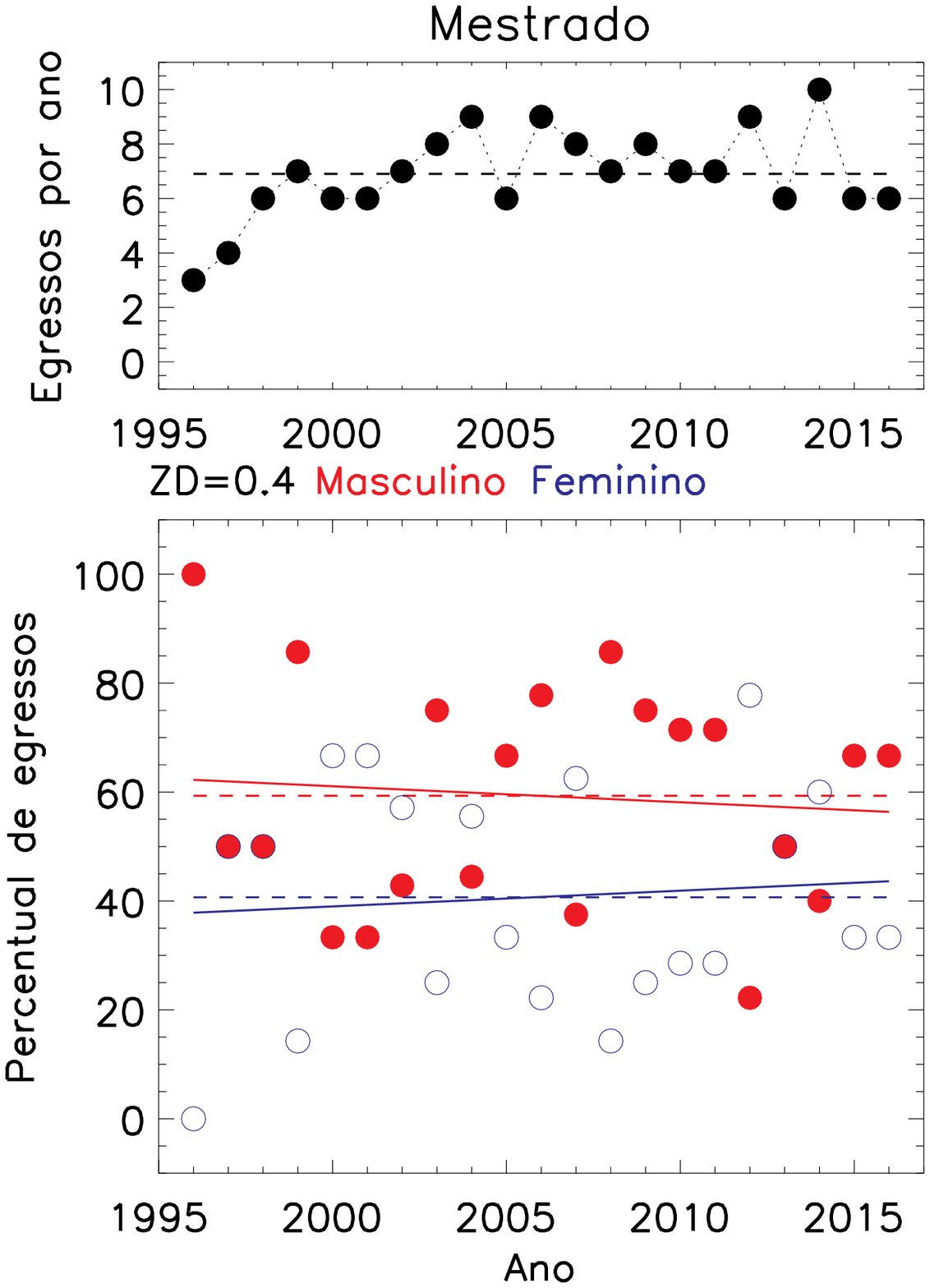}
\includegraphics[width=0.4\textwidth]{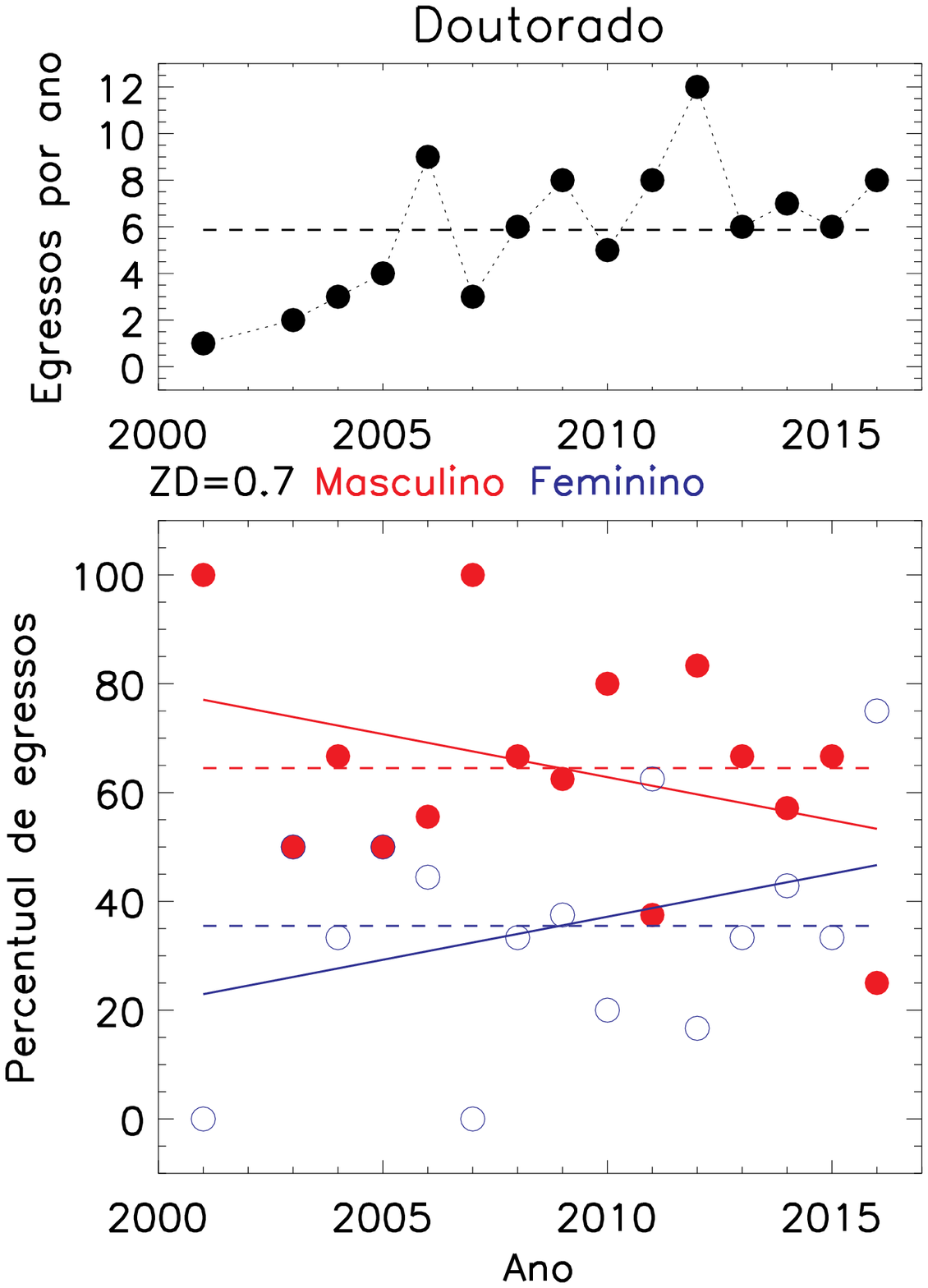}
\caption{Idem a Figura~\ref{graduacao} para os cursos de mestrado (esquerda) e doutorado (direita) em física da UFSM.}
\label{pg}
\end{figure*}

A Figura~\ref{pg} mostra o número de titulados em cada ano nos cursos de mestrado (esquerda) e doutorado (direita) pelo PPG física - UFSM nos paineis superiores, enquanto que os paineis inferiores mostram o percentual de egressos de cada gênero. O número médio de egressos por ano para os dois cursos apresenta um crescimento nos primeiros anos, durante a fase de consolidação do PPG física. Posteriormente, a taxa de egressos tem se mantida aproximadamente constante com 6,9 e 5,9 mestres e doutores titulados por ano, respectivamente. Nos paineis inferiores, observa-se que o percentual de egressos de cada gênero não tem variado significativamente com o passar dos anos, como indicado pelos baixos valores de $ZD$, os quais indicam que não há correlação entre o número de egressos e a data de titulação. Em média, 42\% dos egressos do mestrado são do gênero feminino. Já para o doutroado o  percentual de egressos do gênero feminino é cerca de 39\,\% do número total de egressos.

\subsection{Evolução das notas médias dos egressos}

Utilizamos o programa SIE\footnote{http://www.cpd.ufsm.br/ajuda/sie/} para compilar as notas médias dos egressos dos cursos de graduação em física.  Para o curso de licenciatura diurno observa-se um sutil crescimento das notas das turmas com o passar do tempo, com um valor médio de 7,52$\pm$0.19. As notas médias dos egressos do curso de licenciatura noturno também apresentam um leve crescimento com o tempo, com nota média de todas as turmas 7,50$\pm$0.08. Já as notas médias dos egressos para o curso bacharelado se mantém aproximadamente constante no tempo com média 7,73$\pm$0.07. Observa-se que as notas médias dos três cursos são similares, sendo consistentes entre si, considerando-se os erros padrão associados.

%
%

\section{Discussões}\label{disc}

\begin{table}
\caption{Percentual de egressos do gênero feminino nos cursos de física da UFSM para três épocas: Todos os tempos (desde a formação), últimos 10 anos e últimos 5 anos.}
\label{tab_egressos}
\begin{center}
\begin{tabular}{ c c c c c}
\hline
           Lic. diurno & Lic. Noturno & Bacharelado & Mestrado & Doutorado \\ 
\hline
\multicolumn{5}{c}{Todos os tempos}\\
\hline
        35             &          37            &     28        &   42       &   39        \\
\hline
\hline
\multicolumn{5}{c}{Últimos 10 anos}\\
\hline
       44             &          35            &     29        &   41       &   39        \\

\hline

\hline

\multicolumn{5}{c}{Últimos 5 anos}\\

\hline
       50             &          37            &     42        &   50       &   42        \\
\hline

\end{tabular}
\end{center}
\end{table}

No que se refere a distribuição de egressos por gênero, podemos comparar os resultados para os cursos de física da UFSM com trabalhos disponíveis na literatura. Tanto para os cursos de graduação, quanto de pós-graduação,  encontramos uma tendência de aumento da participação feminina com o passar do tempo, a qual fica mais evidente para o curso de física licenciatura diurno, que apresenta o maior intervalo temporal de dados.  Este resultado está de acordo com trabalhos anteriores \citep{bezerra16,ferrari17,saitovich15} que indicam um sutil aumento da participação feminina na área de física. 

Com o objetivo de comparar o percentual feminino de egressos dos cursos de física da UFSM com outros trabalhos e verificar se houve um aumento recente de mulheres entre os egressos,  na Tabela~\ref{tab_egressos} apresentamos os valores percentuais médios de egressos do gênero feminino nos diferentes cursos de física da UFSM, calculados em três intervalos de tempo distintos: i) desde a formação do curso até 2016; ii) últimos 10 anos e iii) últimos 5 anos.  Todos os valores apresentados na tabela estão acima dos valores médios para o Brasil \citep{saitovich15}, os quais indicam cerca de 25\,\% e 15\,\% de mulheres entre os egressos de cursos de graduação e doutorado em física, respectivamente. 

 Outro resultado em discrepância com as estatísiticas nacionais, refere-se ao fato de que na UFSM não observa-se um decrésmimo da participação feminina no doutorado, em comparação com a graduação. Pelo contrário, os números apresentados na Tab.~\ref{tab_egressos} indicam um aumento da participação feminina nos cursos de pós-graduação. Considerando-se a média dos últimos 5 anos, observa-se que o percentual de mulheres atinge 50\% para os cursos de licenciatura diurno e mestrado em física. No caso do curso de licenciatura, há um claro aumento do número de egressos do gênero feminino com o passar do tempo, conforme indicado na Figura~\ref{graduacao}. Para fim de comparação, a o percentual de mulheres dos egressos nos primeiros 5 anos do curso de licenciatura (1972-1977) foi de cerca de 21\,\%. 

\section{Considerações finais}\label{conc}

Apresentamos uma análise do perfil dos egressos dos cursos de graduação e pós-graduação em física da UFSM, os quais são responsáveis pela formação de  cerca de 500 licenciados, 135 bacharéis, 150 mestres e 90 doutores em física. Os principais resultados são os seguintes:

\begin{itemize}
\item A taxa de egressos dos cursos de graduação tem se mantida aproximadamente constante, sendo de 7,6 titulados/ano para o curso de licenciatura diurno,  7,3 graduados/ano para a licenciatura noturna e 6,5 para o bacharelado. Estes números indicam que menos de 30\,\% dos alunos que ingressam na graduação, a concluem.

\item A taxa de titulados do programa de pós-graduação em física não varia muito, formando 6,9 e 5,9 mestres e doutores por ano, respectivamente.

\item Para todos os cursos, observa-se um aumento do percentual de mulheres entre os egressos com o passar do tempo. O percentual de egressos do gênero feminino é bastante superior à média nacional. Observa-se também um crescimento da participação feminina nos cursos de mestrado e doutorado em física, contrariando a tendência vista em outras instituições.

\item Considerando-se somente os últimos 5 anos, o percentual de egressos do gênero feminino atinge 50\,\% nos cursos de licenciatura diurno e mestrado. Para os demais cursos, o valor percentual médio é de cerca  de 40\,\%.

\item As notas médias dos egressos dos três cursos de física da UFSM se mantiveram constantes com o passar do tempo com um valor médio de aproximadamente 7,5.
\end{itemize}

Observamos que a UFSM apresenta números mais próximos dos desejáveis no que ser refere a questão de gênero nos cursos de física em comparação com a média nacional, porém ressalta-se que o número de egressos do gênero feminino ainda deve aumentar para que se atinja uma situação de equiparidade entre gêneros.
 
Finalmente, este trabalho revela a necessidade da disponibilização pública de dados referentes ao gênero dos egressos por instituições de ensino superior do Brasil, de forma que estudos semelhantes possam ser realizados para egressos de outras instituições. Assim será possível a obtenção de um cenário nacional sobre a participação feminina nos cursos de Física e ciências em geral, o qual poderá ser comparado com resultados de instituições do exterior. 





\bibliographystyle{cen}
\bibliography{refs}

\end{document}